\documentclass[french]{article-hermes}

\usepackage[frenchb]{babel}
\usepackage[latin1]{inputenc}
\usepackage{cite,wrapfig,subfigure,moreverb,verbatimfiles,alltt} 
\usepackage{language,hyperref}

\newif\ifpdf
\ifx\pdfoutput\undefined
    \pdffalse 
\else
    \pdfoutput=1 
    \pdftrue
\fi

\ifpdf
    \usepackage[pdftex]{graphicx}
\else
    \usepackage{graphicx}
\fi

\newcommand{\eng}[1]{\textit{#1}}
\newcommand{\dfn}[1]{\emph{#1}}
\newcommand{\cf}[1]{\textit{cf.} §~\ref{#1}}

\begin{document}

\title[Pandora]{%
Pandora: une plate-forme efficace pour la construction
d'applications autonomes}
\author{Simon Patarin\fup{*} \andauthor Mesaac Makpangou\fup{**}}
\address{\fup{*}Università di Bologna -- Dipartimento di Scienze
  dell'Informazione\\
Mura Anteo Zamboni, 7\\
I-40127 Bologne, Italie\\[3pt]
patarin@cs.unibo.it\\[11pt]
\fup{**}Projet REGAL -- INRIA Rocquencourt\\
Domaine de Voluceau, B.P. 105\\
F-78153 Le Chesnay Cedex, France\\[3pt]
mesaac.makpangou@inria.fr}

\resume{  
  L'informatique autonome a récemment été proposée comme une réponse à
  la difficulté de gérer au quotidien des applications dont la
  complexité ne cesse d'augmenter. Les applications autonomes devront
  être particulièrement flexibles et pouvoir se surveiller en permanence.
  Cette étude présente une plate-forme, Pandora, qui facilite la
  construction d'applications qui satisfont ce double objectif. 
  Pandora s'appuie sur un mode de programmation original des
  applications --- fondé sur la composition de couches et le passage
  de messages --- pour aboutir à un modèle et une architecture
  minimalistes qui lui permettent de contrôler les surcoûts imposés
  par la complète réflexivité de la plate-forme. Un prototype
  fonctionnel de la plate-forme a par ailleurs été développé en C++.
  Une étude détaillée des performances, ainsi que des exemples
  d'utilisation, complètent cette présentation.
}
\abstract{
  Autonomic computing has been proposed recently as a way to address
  the difficult management of applications whose complexity is
  constantly increasing. Autonomous applications will have to be
  especially flexible and be able to monitor themselves
  permanently. This work presents a framework, Pandora, which eases
  the construction of applications that satisfy this double
  goal. Pandora relies on an original application programming pattern
  --- based on stackable layers and message passing --- to obtain
  minimalist model and architecture that allows to control the
  overhead imposed by the full reflexivity of the framework. Besides,
  a prototype of the framework has been implemented in C++. A detailed
  performance study, together with examples of use, complement this
  presentation.
}
\motscles{Informatique autonome, modèle de composants, reconfiguration
  dynamique}

\proceedings{DECOR'2004, Déploiement et (Re)Configuration de Logiciels}{15}

\maketitlepage

\section{Introduction}
\label{sec:introduction}

Les applications distribuées à large échelle occupent une place
grandissante. Les réseaux de distribution de contenu, les grilles de
calcul, les systèmes d'échange de fichiers pair-à-pair, les tables de
hachages distribuées, les systèmes omniprésents: les exemples sont
nombreux et leur liste augmente constamment. L'environnement dans
lequel ces applications sont déployées, Internet, dispose de
caractéristiques originales que sont l'hétérogénéité, l'évolution
rapide de l'ensemble des acteurs (matériels, logiciels, mais aussi
humains) et un relatif manque de fiabilité. La diversité des
plates-formes matérielles et logicielles rend alors particulièrement
difficile la configuration de ces applications. Si l'on parvient
malgré tout à franchir cette première étape, les évolutions et les
possibles pannes peuvent remettre en question les choix faits au
préalable et réduire à néant les efforts précédents. Il est donc
impératif de faciliter ces opérations en les automatisant autant que
possible.  Ces constats sont à la base du développement de
l'informatique autonome (\eng{autonomic computing})\cite{kephart03}
dont le but est d'incorporer aux applications les moyens de
diagnostiquer les difficultés auxquelles elles font face et d'y
remédier elles-mêmes. Les problèmes à résoudre avant de parvenir à
une solution satisfaisante sont évidemment très nombreux. Nous nous
consacrons à l'étude de l'un d'entre eux: le support système
nécessaire au développement de telles applications.

Il est possible d'identifier plusieurs fonctionnalités qui devront
être proposées par une plate-forme de construction d'applications
autonomes. La première et la plus importante concerne la flexibilité
des applications: il est en effet inutile d'imaginer l'autonomie d'une
application si celle-ci ne peut pas être modifiée et reconfigurée
dynamiquement. De nombreux types de reconfiguration sont envisageables
et tous devront être supportés: depuis la simple paramétrisation à
l'ajout de propriétés non fonctionnelles susceptibles de modifier
radicalement le comportement de l'application.  Par ailleurs,
l'application elle-même est souvent la mieux placée pour collecter les
mesures qui lui permettront d'analyser son comportement: il faut donc
que la plate-forme fournisse les mécanismes nécessaires à la diffusion
de ces observations. Cette dernière, enfin, doit faciliter
les interactions entre les différents éléments du système, en
particulier donner accès à l'état courant des applications et aux
mesures effectuées (ce qui exige donc que la plate-forme soit
parfaitement réflexive\cite{smith84,maes87}).

Cette flexibilité que nous requérons ne doit pas être obtenue au
détriment des performances; ce n'est cependant pas l'approche retenue
dans les systèmes actuels où une seule de ces deux propriété est
développée au détriment de l'autre. Ainsi, dans le cas des
plates-formes spécifiquement conçues pour le développement
d'applications autonomes (comme AutoPilot\cite{ribler98} ou
AutoMate\cite{agarwal03}), la flexibilité (et donc les possibilités de
reconfiguration) des applications est mise à mal et est notamment
insuffisante pour répondre à la diversité des besoins que nous
anticipons. À l'inverse, dans le domaine de la programmation par
aspects\cite{kiczales97} ou celui des systèmes à
composants\cite{clarke01}, les applications construites peuvent faire
preuve d'une très grande flexibilité, mais les surcoûts introduits
réduisent considérablement le niveau de performance.

L'approche que nous défendons pour répondre à ce problème repose sur
un compromis original. Plutôt que de transiger sur les possibilités de
flexibilité offertes ou le langage utilisé, nous mettons en avant
l'utilisation d'un mode de programmation alternatif: l'empilage de
composants indépendants. Cette approche, même si elle n'a encore
jamais été employée dans ce contexte à notre connaissance, n'est
cependant pas nouvelle et plusieurs projets l'ont expérimentée,
soulignant son expressivité.  Parmi les premiers travaux dans ce
domaine, on retrouve \textit{x}-Kernel\cite{hutchinson91} et
Ficus\cite{heidemann91} (respectivement un système d'exploitation
spécialisé dans les communications et une méthode de construction de
systèmes de fichiers); plus récemment, deux nouvelles architectures
ont été proposées: un routeur flexible, le \eng{Click Modular
  Router}\cite{kohler00} et SEDA\cite{welsh01} qui permet la
construction efficace de services Internet. Alors que la plupart des
systèmes patrimoniaux doivent instrumenter les appels de procédure (ce
qui se révèle coûteux et pousse à utiliser un langage interprété), ce
mode de programmation, permet de définir le degré d'intervention voulu
au travers du choix de la granularité des composants. La plate-forme
que nous présentons, Pandora, s'appuie donc sur ces techniques et
fournit une interface réflexive qui permet aux applications
extérieures, ainsi qu'aux composants eux-mêmes, de reconfigurer
dynamiquement le système entier.

Nous allons maintenant présenter (section~\ref{sec:art}) un rapide
panorama des travaux sur lesquels s'appuie cette étude. Nous décrivons
ensuite le modèle de composants de Pandora (section~\ref{sec:modele})
et la grande flexibilité de la plate-forme qui en résulte
(section~\ref{sec:flexibilite}). Nous précisons ensuite la mise en
{\oe}uvre de cette plate-forme et donnons quelques exemples
d'utilisation du prototype (section~\ref{sec:implementation}). Nous
terminons cette étude par une évaluation des performances de Pandora
(section~\ref{sec:performance}) et quelques remarques de conclusion
(section~\ref{sec:conclusion}).

\section{Travaux comparables}
\label{sec:art}

Comme nous l'avons mentionné, il existe quelques plates-formes qui
poursuivent des objectifs comparables aux nôtres. Il s'agit par
exemple d'AutoPilot\cite{ribler98} qui se focalise plus
particulièrement sur la conception de moyens d'observation intégrés
aux applications, ou d'AutoMate\cite{agarwal03} qui cible pour sa part
les applications déployées sur les grilles de calcul. La flexibilité
de ces plates-formes est relativement limitée: le paramétrage
dynamique autorise, au mieux, le choix entre différentes mises en
{\oe}uvre pour des fonctionnalités prédéfinies et les possibilités
d'extension et de modification de propriétés non fonctionnelles sont
quasiment inexistantes. Ce manque de flexibilité nous a poussé à
considérer les approches actuellement proposées pour répondre à ce
problème: les systèmes à composants et la programmation orientée
aspects.

Les systèmes à composants patrimoniaux --- .NET, CCM (\textit{Corba
  Component Mo\-del}) ou EJB (\eng{Enterprise Java Beans}) --- sont en
réalité relativement peu adaptés à la conception d'architectures
flexibles. La granularité importante des composants, les liaisons
entre composants difficilement modifiables dynamiquement et l'ensemble
limité et prédéfini de propriétés non fonctionnelles fournies par les
conteneurs de composant y contribuent grandement.  Ceci a favorisé le
développement de systèmes plus légers (et plus performants). Ainsi, la
plate-forme OpenCOM\cite{clarke01} autorise la reconfiguration
dynamique des composants et des liaisons qui existent entre eux par le
biais d'une interface réflexive. Un autre exemple est
Gravity\cite{hall03} qui adopte une démarche originale: le système
établit dynamiquement les liaisons entre les composants selon les
interfaces qu'il fournit et celles qu'il requiert. Pour notre
application la principale limitation d'un tel système est l'absence de
support pour les reconfigurations «simples» comme la modification de
la valeur d'un paramètre.

La programmation orientée aspects\cite{kiczales97} est une
méthodologie qui promeut la séparation des préoccupations
(\eng{separation of concerns}): les fonctionnalités transverses à
plusieurs modules d'un programme sont isolées (ce sont les
\dfn{aspects}) et seront «tissées» avec le reste de l'application au
moment de la compilation ou de l'exécution. La flexibilité de ces
architectures provient de la relative indépendance entre les
différentes entités (modules et aspects) de l'application que cette
approche procure: il est alors possible de modifier un module ou un
aspect sans perturber le reste du programme. C'est la plate-forme
JAC\cite{pawlak02b} (\eng{Java Aspect Components}) qui se rapproche le
plus de nos objectifs. Ici, en effet, les aspects sont encapsulés sous
la forme de composants et tissés à l'exécution (ce qui en autorise
donc la reconfiguration dynamique). La principale limitation de cette
plate-forme est son manque de support pour la reconfiguration des
modules eux-mêmes (ceux dont les fonctionnalités ne peuvent pas être
vues comme des aspects).

\section{Modèle de composants}
\label{sec:modele}

Pandora s'appuie sur la notion de \dfn{composant} comme unité de
travail élémentaire et indépendante; ces composants peuvent être
assemblés sous forme de piles pour exécuter des tâches de plus haut
niveau. Nous allons décrire plus précisément le modèle de composant
qu'utilise Pandora et nous présentons ensuite --- brièvement --- le
langage de description d'architecture utilisé pour définir et
paramétrer ces assemblages de composants.

Contrairement à la plupart des modèles de composants existants, les
composants utilisés par Pandora communiquent exclusivement par passage
de message et non par invocation. Cette contrainte contribue
grandement à réduire la complexité des composants: un composant ne
définit en effet qu'une seule interface, celle qui lui permet de
recevoir un message. Dans la terminologie utilisée, les messages
échangés entre les composants sont appelés événements, et c'est ce
terme que nous emploierons par la suite.  Les événements sont typés et
un composant a la possibilité d'utiliser cette information pour
déterminer les traitements qu'il aura à effectuer.

Chaque composant dispose, au maximum, d'un port d'entrée et d'un
nombre quelconque de ports de sortie. Le transfert d'événements entre
les composants est unidirectionnel, synchrone et se fait par flot
continu.  Ceci signifie, en particulier, que deux composants, une fois
la communication établie, sont durablement associés l'un à l'autre et
s'oppose à la transmission de messages par \dfn{boîte aux lettres}
(nous verrons plus loin que ce mode de transmission existe néanmoins
dans le système). L'ensemble des composants associés par de telles
relations est appelé \dfn{pile}. Un nom est associé à chaque pile et
il est possible de faire coexister plusieurs instances d'une même pile
dans le système qu'on peut différencier au moyen de «poignées»
(\eng{handles}) attribuées par le système ou bien d'\dfn{alias}
spécifiés explicitement.

Du fait du synchronisme de la transmission des événements au sein
d'une pile, les traitements des composants sont directement associés à
la réception d'un événement, lesquels peuvent à leur tour entraîner la
transmission d'un ou plusieurs événements. Ainsi, une fois qu'un
événement est «produit» l'ensemble des traitements générés (par
transmission successive d'événements) est traité séquentiellement. La
production initiale des événements est assurée dans chaque pile par un
composant spécifique nommé \dfn{composant initial} qui est ordonnancé
de manière indépendante.  Ce composant --- il n'en existe qu'un seul
par pile --- est donc indirectement responsable de l'activité de
l'ensemble de la pile: s'il cesse de produire des événements, c'est
toute la pile qui est au repos pendant ce temps.

Un autre caractéristique importante des composants est leur entière
indépendance: un composant ignore totalement le contexte dans lequel
il sera utilisé. Ainsi les ports d'entrée et de sortie sont
parfaitement anonymes et un composant n'a pas la possibilité de
choisir le type des composants avec lesquels il communique; ceci est
déterminé au moment de la configuration de la pile. Dans le cas
général, un composant qui désire transmettre un événement le transmet
donc à son \dfn{successeur}, sans connaître son identité. Lorsqu'un
composant dispose de plusieurs ports de sortie, il existe deux
possibilités (à l'exclusion l'une de l'autre):
\begin{enumerate}
\item \emph{alternative}: des ports alternatifs sont identifiés par un
  numéro d'ordre et il sera possible de configurer la pile pour que
  des composants de nature différente correspondent à chacun des
  ports;
\item \emph{démultiplexage}: des ports de démultiplexage permettent à
  un composant de classifier les événements: dès qu'une nouvelle
  catégorie est découverte, un nouveau port est créé dynamiquement,
  alors que les événements appartenant à une catégorie existante sont
  dirigés vers le port qui y avait été précédemment associé.
\end{enumerate}
Nous voyons alors que chaque composant à sorties multiples doit être
configuré à l'aide de «sous-piles» pour chaque port de sortie. Ces
sous-piles sont nommées \dfn{branches} et correspondent aux
traitements différenciés des événements au sortir d'un composant à
sorties multiples. Ainsi, tous les événements qui sortent des branches
associés à un composant à sortie multiple donné sont envoyés sur le
port d'entrée d'un unique composant qui les multiplexe.

Chaque composant, outre les ports de sortie que nous avons mentionné,
a la possibilité de communiquer directement avec une pile, prise dans
son ensemble; le port d'entrée d'une pile est alors défini par le port
d'entrée de son composant initial. À la différence des communications
entre composants qui sont anonymes, les communications entre piles
sont nommées: un composant indique le nom de la pile (ou son
\eng{alias}) à laquelle il souhaite transmettre des événements. Ici,
les communications peuvent être, au choix, synchrones ou asynchrones:
ce modèle de communication peut donc être utilisé pour des
transmissions d'événements selon le modèle des boîtes aux lettres.
Les composant, enfin, ont la possibilité de déclarer explicitement des
paramètres, que nous nommons \dfn{options}, identifiés par un nom et
dont la valeur peut être configurée pour adapter leur comportement.
Ces options peuvent être de tout type et il est possible d'associer un
traitement particulier à l'affectation d'une valeur (par exemple
transformer une chaîne de caractères représentant un nom de machine en
adresse IP numérique).

La spécification de l'assemblage des composants au sein des piles,
ainsi que leur paramétrisation initiale, se fait au moyen d'un langage
spécifique. Afin de faciliter son utilisation et sa compréhension par
les utilisateurs du système, une représentation compacte des piles
nous a paru préférable. Il est cependant envisageable d'utiliser des
langages de description de graphe (tel que
\texttt{dot}\cite{ellson02}) ou des langages à balises (de type XML)
pour aboutir à des résultats similaires. Nous nous limiterons ici à
présenter très brièvement la grammaire de ce langage
(tableau~\ref{tab:stack_grammar}) et un exemple complet d'utilisation
(figure~\ref{fig:stack_example}), sa compréhension ne posant pas de
problème particulier (voir\cite{patarin03fr}, section 3.4.4 pour une
description détaillée).

\begin{table}[htbp]
  \centering
{\small\begin{tabular}{l@{~~::~~}l}
\hline
\multicolumn{2}{c}{} \\
\texttt{pile} &      \texttt{`\%'id alias? `\{' composant* `\}'} \\
\texttt{composant} & \texttt{simple | demux | alternative} \\ 
\texttt{simple} &    \texttt{`@'id alias? options?} \\
\texttt{options} &   \texttt{(`[' option (`,' option)* `]')?} \\
\texttt{alias} &     \texttt{`:'id} \\
\texttt{demux} &     \texttt{simple `<' branche `>}' \\
\texttt{alternative} &    \texttt{simple `(' branche (`|' branche)* `)'} \\
\texttt{branche} &   \texttt{composant+} \\
\texttt{option} &    \texttt{`\$'id alias? (`=' valeur)?} \\
\texttt{id} &        \texttt{[a-zA-Z]([a-zA-Z0-9\_])*} \\
\texttt{valeur} &    \texttt{entier | flottant | booléen | `"'caractères`"'} \\
\multicolumn{2}{c}{} \\
\hline
  \end{tabular}}
  \caption{Grammaire du langage de description d'architecture (BNF).}
  \label{tab:stack_grammar}
\end{table}

\begin{figure}[htbp]
  \centering
  {\small \verbatimlisting{stack_example.txt}}
  \caption{Exemple complet de la configuration d'une pile (utilisée pour
  la capture et la reconstruction du trafic DNS).}
  \label{fig:stack_example}
\end{figure}

\section{Plate-forme flexible}
\label{sec:flexibilite}

Nous avons dit à quel point les applications autonomes avaient
besoin d'être fle\-xi\-bles. Ceci s'exprime à la fois dans la capacité à
les paramétrer le plus finement possible, mais également dans la
possibilité de revenir sur ces choix au fur et à mesure que le
contexte dans lequel s'exécutent les applications évolue. Plutôt que
de décrire l'architecture dans son ensemble (décrite avec précision
dans des travaux précédents\cite{patarin03fr}), nous allons nous
concentrer sur ce point précis et détailler les mécanismes mis à
disposition pour le développement d'applications autonomes.

\subsection{Degrés de flexibilité}

Deux niveaux de flexibilité apparaissent dans la spécification des
systèmes manipulés par Pandora: au niveau du paramétrage des
composants et au niveau de l'assemblage des composants. La flexibilité
liée au paramétrage passe par l'utilisation des options et est tout à
fait classique. Elle peut cependant avoir une influence importante
puisqu'il est par exemple envisageable d'utiliser des options pour
spécifier un algorithme de démultiplexage ou bien un port de sortie au
sein d'une alternative.

Pandora expose un second niveau de flexibilité dans la spécification
des assemblages de composants. D'un côté, la possibilité d'avoir
différents composants qui mettent en {\oe}uvre la même fonctionnalité
permet de choisir la solution qui correspond le mieux à
l'environnement dans lequel on se trouve (les algorithmes pour
lesquels il faut trouver un compromis entre utilisation du processeur
et occupation mémoire ne sont pas rares). D'un autre côté, le modèle
autorise l'utilisation de composants non fonctionnels dont l'insertion
dans une pile n'en modifie le comportement général mais peut altérer
la façon dont les traitements sont exécutés ou y adjoindre certains
effets de bord. Ainsi, on peut citer l'exemple de composants assurant
la persistance des événements qu'ils reçoivent, la répartition des
traitements sur différentes machines, la surveillance de
l'application, la détection des pannes, la synchronisation des
traitements, etc. Ce dernier point se rapproche des problématiques que
l'on retrouve dans la programmation par aspects: sans modifier
l'application (les composants orignaux), il est possible d'y tisser
certains aspects (par l'adjonction de composants spécifiques dans la
définition de la pile).

\subsection{Reconfiguration dynamique}
\label{sec:reconfiguration}

L'ensemble de la configuration de la plate-forme, ainsi que son état
courant, sont exposés par son moteur d'exécution (qui prend la forme
d'un micro-noyau) au travers d'une interface de réflection. Ceci
concerne donc aussi bien les définitions de piles, que les valeurs des
options ou les listes de ressources: absolument tout ce qui est
configurable par un fichier de configuration l'est également au
travers de cette interface. Pour chaque élément, il est en outre
possible de préciser si l'on souhaite modifier les définitions
stockées ou leur représentation active (modifiant ainsi directement le
comportement de la plate-forme). La gestion des piles est également
exposée, de telle sorte qu'il est possible de connaître les piles en
cours d'exécution ou de demander le démarrage ou l'arrêt de l'une
d'entre elles.

Parmi ces opérations, la reconfiguration dynamique des piles occupe
une place particulière. En effet, contrairement à toutes les autres,
cette manipulation conduit à modifier des instances de composant
existantes ainsi que les liaisons qui existent entre elles. Les
composants de Pandora sont considérés par défaut comme porteur d'un
état et la plate-forme veille donc à réduire au strict nécessaire les
destructions de composants lorsque l'on passe de l'ancienne
description à la nouvelle.

\subsection{Modes de contrôle}

Ces différentes opérations sont rendues accessibles pour les
applications externes à la plate-forme au travers d'un protocole à
meta-objets (\eng{Meta-Object Protocol}). Pandora fournit une
interface à ce protocole spécifique dans différents langages de
programmation, notamment C, C++, Perl et Guile\cite{guile}. Il est
ainsi possible d'écrire de véritables «scripts de contrôle», ce qui
favorise le développement rapide d'applications autonomes.

Nous souhaitons également que les applications puissent s'analyser et
se reconfigurer elles-mêmes. L'utilisation du protocole que nous
venons de décrire n'est alors pas optimale en termes de performance,
nous avons donc introduit un mécanisme spécifique de \dfn{capteurs} et
de \dfn{moniteurs} pour répondre à ce problème. Cette technique permet
à un objet (le moniteur) d'appliquer des traitements sur un ensemble
de valeurs (exposées par les capteurs) de manière efficace.  Un schéma
de nommage simplifié permet en effet de réduire considérablement les
surcoûts qui sont liés dans l'approche précédente à la localisation
des paramètres au sein du système (à chaque accès il faut en effet
identifier la pile, puis le composant dans la pile et enfin le
paramètre dans le composant). De plus, les capteurs supportent
plusieurs modes de fonctionnement: un mode \dfn{passif} où les
moniteurs décident eux-mêmes de procéder aux traitements et un mode
\dfn{actif} où la modification de la valeur d'un capteur déclenche les
traitements.

\section{Mise en {\oe}uvre}
\label{sec:implementation}

Nous avons développé une plate-forme logicielle qui met en {\oe}uvre
cette architecture. Elle représente environ 50\,000 lignes de code en
C++ et comprend, outre le noyau, plus d'une centaine de composants.
Environ le tiers de ces composants fait partie des composants «de
base»: ce sont des composants qui mettent en {\oe}uvre des propriétés
non fonctionnelles et peuvent être utilisés dans toute pile.

Pandora est utilisable sur un grand nombre de systèmes dont
Linux, FreeBSD, NetBSD, Solaris, Digital Unix (Tru64) et, pour le
noyau et les composants de base uniquement, Win32. La plate-forme est
distribuée dans sa version la plus récente sous une licence
\eng{open-source} de l'INRIA (libre pour une utilisation non
commerciale) à l'adresse suivante:
\url{http://www-sor.inria.fr/projects/relais/pandora/}.

La plate-forme a déjà été utilisée dans plusieurs
projets\cite{patarin03fr, patarin03a, lefessant03, patarin04a}.
L'application néanmoins qui met le mieux en valeur la flexibilité de
Pandora et les possibilités qu'elle offre pour construire des
applications autonomes est C/SPAN\cite{ogel03}. C/SPAN est un cache
Web autonome qui s'appuie d'une part sur C/NN\cite{ogel01}, un cache
flexible, et sur une pile d'observation du trafic HTTP au-dessus de
Pandora d'autre part. Dans ce système, C/NN et Pandora forment une
boucle d'interaction rapprochée: C/NN, en fonction de son
environnement (espace disque, taux de requêtes, etc.) ajuste le
comportement de Pandora en utilisant son interface réflexive, tandis
que Pandora reconfigure le cache en fonction de ses observations.

\section{Évaluation des performances}
\label{sec:performance}

\begin{figure}
  \centering
  \subfigure[Comparaison des temps d'exécution de diverses
  fonctions avec le temps de traversée d'un composant.]{%
    \includegraphics[width=.49\hsize]{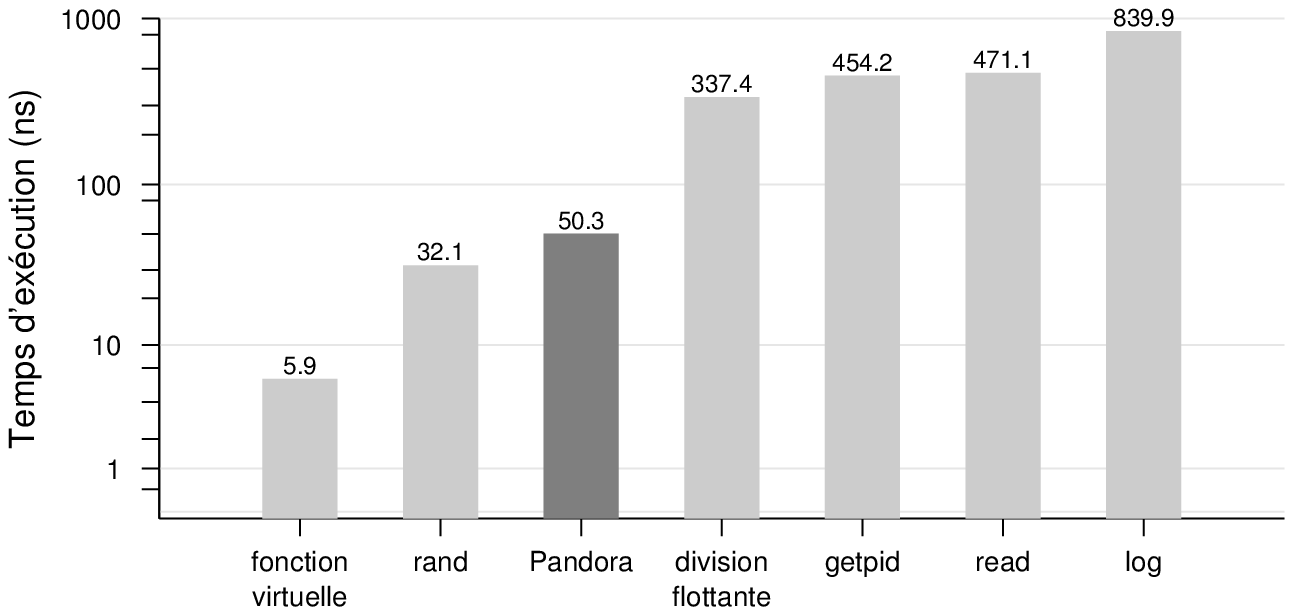} 
    \label{fig:perf_compcall}}\hfill 
  \subfigure[Comparaison des mécanismes d'introspection de Pandora
  avec ceux fournis en standard avec Java et C\#.]{%
    \includegraphics[width=.49\hsize]{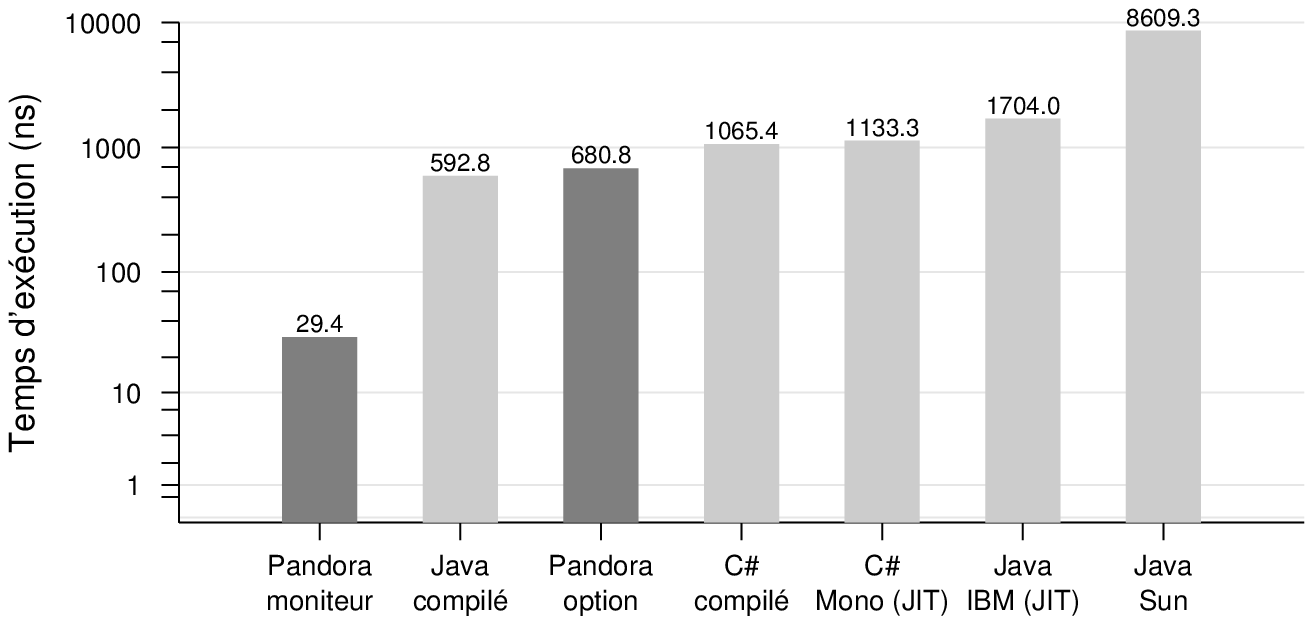}
    \label{fig:perf_reflect}}
  \caption{Évaluation des performances}
  \label{fig:perf}
\end{figure}

Dans cette évaluation des performances de la plate-forme, nous nous
sommes concentrés sur deux points particuliers: le coût du découpage
en composants et celui des opérations d'introspection. 

L'ensemble de ces tests a été réalisé sur une même machine munie d'un
Pentium~IV cadencé à 2,4~GHz qui exécute la version 2.6 du système
d'exploitation Linux. Les mesures que nous présentons ici sont
calculées à partir de la moyenne des valeurs obtenues pour 50
évaluations successives; à aucun moment l'erreur standard associée à
ces moyennes n'a dépassé 1\,\%. Les temps d'exécution des procédures
ont été calculés à partir d'une boucle qui les répète un million de
fois chacune: le temps total exprimé en millisecondes indique alors le
coût d'une itération, exprimé en nanosecondes.

Pour évaluer le coût du découpage en composants, nous avons mesuré le
temps de traversée d'un composant, c'est-à-dire le temps nécessaire à
la transmission d'un événement d'un composant à son successeur. Les
résultats présentés figure~\ref{fig:perf_compcall} montrent que ce
temps est d'environ 50\,ns. D'après les autres mesures que nous avons
effectuées, nous voyons que ce temps est certes supérieur au temps
nécessaire à l'appel de fonctions de bibliothèques simples mais est
inférieur, par exemple, à celui utilisé pour effectuer une opération
flottante ou un appel système. Ceci nous indique donc que pour des
applications non triviales, le surcoût lié au découpage en composants
reste très limité, voire négligeable.

Pour surveiller son propre comportement, une application autonome
examine en permanence les capteurs dont elle dispose. Les modifications
ne sont en effet supposées intervenir que lors de circonstances
exceptionnelles. C'est donc l'opération de lecture d'un capteur qui
est la plus critique en termes de performance, et c'est elle que nous
avons donc choisi d'évaluer. À titre de comparaison, nous avons mesuré
le temps nécessaire à la lecture d'une variable en utilisant les
interfaces réflexives de deux langages couramment utilisés dans la
construction d'applications autonomes: Java et C\#.  L'opération
effectuée consiste ici à lire la valeur d'un champ entier d'un objet
de la manière la plus simple (le code tient, dans les deux cas, en une
seule ligne). Pour Java nous avons utilisé différentes machines
virtuelles, avec et sans compilation dynamique (\eng{Just In Time});
nous avons également compilé le programme en code natif en nous
servant du compilateur GNU\cite{gcj}. Pour C\#, c'est l'environnement
Mono\cite{mono} que nous avons employé, dont la machine virtuelle
supporte à la fois la compilation dynamique et statique (avant
l'exécution). Les résultats de ces différents tests sont présentés
figure~\ref{fig:perf_reflect}.  Plutôt que de nous étendre sur les
performances relatives des compilateurs ou des machines virtuelles les
uns par rapport aux autres, ce sont les ordres de grandeurs qui nous
intéressent ici. Il apparaît que l'utilisation des capteurs --- en
mode passif --- de Pandora (\cf{sec:controle}) est environ 20 fois
plus rapide que le code Java compilé. Le temps d'exécution de ce
dernier correspond en revanche à celui requis lorsque l'on utilise les
options de Pandora au travers de l'interface externe. L'utilisation
d'une machine virtuelle dégrade encore un peu les performances et
l'absence de compilation dynamique les rend tout à fait
catastrophiques. Ceci montre donc que l'utilisation de Pandora permet,
à charge égale, de supporter simultanément un nombre bien supérieur
(un ou deux ordres de grandeur) de capteurs et donc d'applications,
comparés aux langages traditionnellement utilisés dans ce domaine.
C'est la spécialisation de Pandora dans ces tâches (la plate-forme a
été spécialement conçue et optimisée pour cela) par opposition à la
nécessaire généricité d'un langage de programmation qui explique ces
différences.

\section{Conclusion}
\label{sec:conclusion}

Nous avons présenté Pandora, une plate-forme pour la construction
d'applications autonomes. Pandora s'appuie sur un mode de
programmation original, l'empilage de composants, qui offre un bon
compromis entre flexibilité et performance. En effet, le modèle de
composants de Pandora qui en résulte est nettement plus simple que les
approches les plus couramment utilisées qui procèdent par appel de
procédure. Le langage de description d'architecture que nous avons
conçu permet de définir les assemblages de composants de façon
compacte et claire pour les utilisateurs de la plate-forme.
L'architecture du système est organisée autour d'un micro-noyau qui
expose une interface réflexive et propose les abstractions nécessaires
au programmeur d'application pour permettre la configuration et
reconfiguration du système tout entier. Cette architecture a été mise
en {\oe}uvre au travers d'un prototype fonctionnel --- disponible
librement sur Internet --- qui a déjà eu de nombreuses applications.
Enfin, une évaluation des performances du système a montré que sa mise
en {\oe}uvre soutenait notre objectif initial de concilier flexibilité
et performance.

\let\tilda\~

\bibliography{decor04}

\end{document}